\documentclass[a4paper,11pt]{article}
\pdfoutput=1 % if your are submitting a pdflatex (i.e. if you have
\usepackage{jheppub} % for details on the use of the package, please
\usepackage{latexsym,amssymb,amstext,amsmath}
\usepackage{hyperref}
\usepackage{graphicx} 
\usepackage{subcaption}
\usepackage{amsthm}
\usepackage{esint}
\usepackage{color}
\begin{document}
%%%%%%%%%%%%%%%%%%%%%%%%%%%%%%%%%%%%%%%%%%%%%%%%%%%%%%%%%%%%%%
%
\begin{titlepage}

\begin{center}
 {\LARGE\bfseries Generating new $N=2$ small black holes}
  \\[10mm]

\textbf{Davide Polini}

\vskip 6mm
{\em  
  Department of Mathematics, 
  Instituto Superior T\'ecnico,\\ Universidade de Lisboa,
  Av. Rovisco Pais, 1049-001 Lisboa, Portugal}\\[.5ex]

{\tt  davide.polini@tecnico.ulisboa.pt }
\end{center}

\vskip .2in
%%%%%%%%%%%%%%%%%%%%%%%%%%%%%%%%%%%%%%%%%%%%%%%%%%%%%
\begin{center} {\bf ABSTRACT } \end{center}
\begin{quotation}\noindent

\vskip 3mm
\noindent
We use the exact symmetries of the $N=2$ STU model of Sen and Vafa to classify BPS orbits in this theory.
Subsequently, we construct examples of small BPS black holes in this model
by solving the associated attractor equations, and use duality symmetries as a solution generating technique to construct a two-center
bound state of small BPS black holes in asymptotically flat space-time.

%%%%%%%%%%%
\end{quotation}
\vfill

\vfill
%%%%%%%%%%
\today
%%%%%%%%%%%%%%%%%%
\end{titlepage}
%%%%%%%%%%%%%%%%%%%%%%%%%%%%%%%%%%%%%%%%%%%%%%%%%%%%%%%%%%%

\tableofcontents

%%%%%%%%
\section{Introduction }
%%%%%%%%%%
The search for a statistical understanding of black hole entropy as well as attempts to extend the solution space of black hole backgrounds in 
various theories of gravity has yielded rich insights into the mathematical structures that govern the non-perturbative physics in these theories. In particular, the former endeavour has
uncovered modular forms that underlie the microstate organisation of certain classes of
supersymmetric dyonic black holes in specific four-dimensional $N=4$ and $N=8$ string theories 
\cite{Dijkgraaf:1996it,Shih:2005uc,Shih:2005qf,Gaiotto:2005hc,Jatkar:2005bh,David:2006ji,Dabholkar:2006xa,
David:2006yn,David:2006ru,David:2006ud,Dabholkar:2006bj,Sen:2007vb,Dabholkar:2007vk,
Banerjee:2007ub,Sen:2007pg,Banerjee:2008pv,Banerjee:2008pu,Sen:2008ta,Dabholkar:2008zy}.
Especially in the $N=4$ cases, where the counting formulae for a 
class of dyonic black holes can be written down exactly in terms of Siegel modular forms, studying the solution space of decadent black hole backgrounds restricted to 2-centered small
black hole pairs and the lines of marginal stability across which they decay 
allows one to extract non-perturbative information encoded in the zeros of these modular forms. 
In both the quest to write down exact counting formulae as well as study the solution 
space of small black holes, the underlying duality symmetries of the theory not only provide useful checks on allowed solutions and modular forms, but also act as solution generating 
techniques in the low energy gravity theories obtained from the string theories under consideration. 
So far,  both types of aforementioned analyses have proved elusive in the $N=2$ case until very recently, when an approximate counting formula, in terms of Siegel modular forms, 
Jacobi and quasi-modular forms was written down \cite{Cardoso:2019avb} for supersymmetric dyonic black holes in 
the $N = 2$ STU model of Sen and Vafa (example D of \cite{Sen:1995ff}).
This is a model that possesses
exact duality symmetries. These have been used recently \cite{CdWM} to obtain exact results for
the function $F$ that encodes the Wilsonian action describing the coupling of vector multiplets to supergravity in the presence of an infinite set of gravitational coupling functions
$\omega^{(n+1)}(S,T,U) $ ($n \geq 0$). The function $F$ is one of the ingredients
that enter in the definition of the quantum entropy function \cite{Sen:2008yk,Sen:2008vm,Dabholkar:2010uh}
for BPS black holes in $N=2$ theories.
In the $N=2$ Sen-Vafa STU model, it is this quantum entropy function for large dyonic BPS black holes that was rewritten in 
\cite{Cardoso:2019avb} as an approximate counting formula, in terms of modular forms.

In this note, we will attempt to harness the exact symmetries of the model to investigate the solution space of 2-centered BPS small black holes. 
We will begin with a classification of BPS orbits using the $\Gamma_0(2)$ duality symmetry of
this $N = 2$ model, following the analysis based on $SL(2)$ given in \cite{Nampuri:2007gw}. We will then turn
to the construction of small BPS black holes in this $N = 2$ model. This is done by solving
the associated attractor equations. To this end, we begin by working in a regime where
the gravitational coupling functions 
$\omega^{(n+1)}(S,T,U) $ simplify, namely, we take
 two of the moduli $S, T, U$ of the model to be small, while the third modulus
is taken to be large. In this regime, the small BPS black holes \cite{Sen:1995in,Dabholkar:2004yr,Dabholkar:2004dq} that we obtain carry more than two charges, in agreement with 
\cite{Cardoso:2008fr}.  
Subsequently, with the help of duality symmetries, we construct a two-center bound state of small BPS
black holes in asymptotically flat space-time, by taking the asymptotic moduli to equal
the attractor values of a dyonic BPS black hole carrying the same amount of charge as the
combined two-center configuration.

%%%%%
\section{Classification of orbits}
%%%%%%

In $N=2$ supergravity theories in four dimensions, the couplings
of Abelian vector multiplets to supergravity are encoded
 in a holomorphic function
$F(Y, \Upsilon)$ that characterizes the Wilsonian effective action \cite{deWit:1996gjy}.
The function $F(Y, \Upsilon)$ can be decomposed as $F(Y, \Upsilon) = F^{(0)} (Y) 
+ 2 i   \, \Omega (Y, \Upsilon)$, where $\Upsilon$ denotes the (rescaled) lowest component of the square of the Weyl  superfield,
and where $Y^I \in \mathbb{C}$ denote complex scalar fields that reside in the off-shell vector multiplets of the underlying
supergravity theory ($I=0, 1, \dots, n$).
These theories admit BPS black holes, which are extremal dyonic black holes carrying 
electric/magnetic charges $(q_I, p^I)$. 

In this note, we will consider small BPS black holes in the $N=2$ STU model
of Sen and Vafa (example D of \cite{Sen:1995ff}). For this model, $I =0,1,2,3$. 
We introduce projective coordinates 
\begin{equation}
S = -i \frac{Y^1}{Y^0} \;\;\;,\;\;\; T= -i \frac{Y^2}{Y^0} \;\;\;,\;\; U = -i \frac{Y^3}{Y^0}\;.
\end{equation}
This model possesses duality symmetries, namely $\Gamma_0(2)_S \times \Gamma_0(2)_T
\times \Gamma_0(2)_U$ symmetry as well as
triality symmetry under exchange of $S, T, U$. Here, $\Gamma_0(2)$ denotes the subgroup
\begin{eqnarray}
\Gamma_0 (2) = \left\{ 
\begin{pmatrix}
a & b\\
c & d\\
\end{pmatrix} \in SL(2, \mathbb{Z}): 
\begin{pmatrix}
a & b\\
c & d\\
\end{pmatrix} \equiv 
\begin{pmatrix}
1 & *\\
0 & 1 \\
\end{pmatrix} \mod 2
\right\} \;,
\label{gam02}
\end{eqnarray}
where $*$ can take any value in $\mathbb{Z}$.

We may assemble the eight 
electric/magnetic charges $(q_I, p^I)$ into two vectors, 
\begin{eqnarray}\label{chargenotation}
Q^T_e &=&  (q_0, - p^1, q_2, q_3) \;, \nonumber\\
Q^T_m &=& (q_1, p^0, p^3, p^2) \;.
\end{eqnarray}
Then, under $\Gamma_0(2)_S$-transformations, these two vectors transform as \cite{Cardoso:2008fr},
\begin{eqnarray}\label{QeQmtrasf}
Q^T_e &\rightarrow a \, Q^T_e - b \, Q^T_m \;, \nonumber\\
Q^T_m &\rightarrow d \, Q^T_m - c \, Q^T_e \;, 
\end{eqnarray}
where
\begin{eqnarray}
\begin{pmatrix}
a & b\\
c & d\\
\end{pmatrix}
\in
\Gamma_0 (2)  \;.
\label{gam02}
\end{eqnarray}

Each of the four-component vectors $Q_e, Q_m$ corresponds to a point in a four-dimensional lattice made out of
two two-dimensional hyperbolic lattices, each 
with hyperbolic metric $h$, 
\begin{eqnarray}
h = \begin{pmatrix}
0 & \, 1 \\
1 & \, 0 
\end{pmatrix} \;.
\end{eqnarray}
Introducing the four-dimensional metric $\eta = h \oplus h$,
we define the charge bilinears
\begin{eqnarray}\label{chargebilinear}
2 n &=& Q_e^T \, \eta \, Q_e = 2 \left( - q_0 p^1 + q_2 q_3 \right) \;, \nonumber\\
2 m &=&  Q_m^T \, \eta \, Q_m 
=
2 \left( p^0 q_1 + p^2 p^3 \right) \;, \nonumber\\
l &=&  Q_e^T \, \eta \, Q_m  = q_0 p^0 - q_1 p^1 + q_2 p^2 + q_3 p^3 \;.
\label{bilin}
\end{eqnarray}
These bilinears transform as a triplet under $\Gamma_0(2)_S$ \cite{Cardoso:2008fr},
\begin{eqnarray}
\begin{pmatrix}
n\\
m\\
l\\
\end{pmatrix} \rightarrow 
\begin{pmatrix}
a^2 & \; b^2 & \;  - ab\\
c^2 & \; d^2 & \;  - cd \\
- 2 ac &\;  - 2 bd &\;  ad + bc 
\end{pmatrix} \begin{pmatrix}
n\\
m\\
l\\
\end{pmatrix} \;,
\end{eqnarray}
and the $\Gamma_0(2)$ invariant norm of this vector is $ \Delta \equiv 4 nm - l^2$.

Given a generic charge vector  $Q_m$, we now show that $Q_m$ can be brought into
one of the following inequivalent forms by means of sequences of $\Gamma_0(2)$-transformations:\\

\noindent
{\bf Proposition:}
Let $Q^T_m =  (q_1, p^0, p^3, p^2) \in \mathbb{Z}^4$ be a charge vector 
with $\gcd  ( q_1, p^0, p^3, p^2) =1$. Then, it can be brought into one of the following
inequivalent forms by $\Gamma_0(2)$-transformations, 
\begin{eqnarray}\label{latticereshape}
Q^T_m &=& (1,  m, 0, 0)  \;\;\;,\;\;\;  m \in \mathbb{Z} \;.
 \nonumber\\
Q^T_m &=& ( m,  1, 0, 0)  \;\;\;,\;\;\;  m \in \mathbb{Z} \;.
 \nonumber\\
Q^T_m &=& (0, 0,  m,  1)  \;\;\;,\;\;\;  m \in \mathbb{Z} \;.
 \nonumber\\
Q^T_m &=& (0, 0,  1,  m) \;\;\;,\;\;\; m \in \mathbb{Z} \;.
\end{eqnarray}
where $m$ is the magnetic charge invariant defined in \eqref{chargebilinear}

\begin{proof}

Given an integral column vector, $ V = \begin{pmatrix} a \\ b \end{pmatrix}$, with non-vanishing entries $a$ and $b$,
we write it as  $ V = g_V \, \begin{pmatrix} \alpha \\ \beta \end{pmatrix}$, 
where $(\alpha, \beta)$ are coprime, and where $g_V = \gcd(|a|,|b|)$
denotes the greatest common divisor of $|a|$ and $|b|$. We then focus
on the vector  $ \begin{pmatrix} \alpha \\ \beta \end{pmatrix}$.
First, consider the case when $\beta$ is even, in which case $\alpha$ is odd.
Then, by a sequence of $T$, $T^{-1}$, $S$ and $S^{-1}$ transformations, this vector can be brought to the
form $ \begin{pmatrix} 1 \\ 0 \end{pmatrix}$.
Here, $T$ and $S$ denote the two generators of $\Gamma_0(2)$, given by 
\begin{eqnarray}
T = 
 \begin{pmatrix} 1  &  1  \\ 0  & 1  \end{pmatrix} 
 \;\;\;,\;\;\;
 S =  \begin{pmatrix}1  &  0  \\ 2   & 1 \end{pmatrix} 
\;.
\label{STgen}
\end{eqnarray}
The $\Gamma_0(2)$-transformation that brings the vector into the form
$ \begin{pmatrix} 1 \\ 0 \end{pmatrix}$ must be of the form
\begin{eqnarray}
 \begin{pmatrix} \delta &  - \gamma \\ -\beta & \alpha \end{pmatrix} 
 \begin{pmatrix} \alpha \\ \beta \end{pmatrix} = \begin{pmatrix} 1 \\ 0 \end{pmatrix}\;,
 \label{trafo1}
\end{eqnarray}
where the integers $(\gamma, \delta)$ are coprime and satisfy
$\alpha \delta - \beta \gamma = 1$,
with $\delta$ odd and $\beta$ even. The fact that $(\gamma, \delta)$ are coprime is a consequence of
the fact that the determinant of any $SL(2, \mathbb{Z})$-matrix is one. The entries of the second row of the
$\Gamma_0(2)$-matrix have to be equal to $(- \beta, \alpha)$ (up to an overall minus sign). The reason is as follows.
Suppose that the entries of the second row are given by coprime integers
$(- \tilde{\beta}, \tilde{\alpha})$, with $ \tilde{\beta}$ even (due to 
the $\Gamma_0(2)$ nature of the matrix), such that $\tilde{\beta} \alpha = \tilde{\alpha} \beta$.
Decompose $\alpha$ into prime factors, $\alpha = p_1 \dots p_n$. Then, none of these prime numbers
can divide $\beta$, since $(\alpha, \beta)$ are coprime. Hence, they must divide $\tilde \alpha$,
and hence $\tilde{\alpha} = m \alpha$, with $m \in \mathbb{Z}$. It follows that $\tilde{\beta} = m \beta$.
Since $(\tilde{\alpha}, \tilde{\beta}  ) $ are coprime, it follows that $m = \pm 1$.

Next, consider the case when
$\beta$ is odd.
Then, by a sequence of $T$, $T^{-1}$, $S$ and $S^{-1}$ transformations, this vector can be brought to the
form $ \begin{pmatrix} 0 \\ 1 \end{pmatrix}$.
The $\Gamma_0(2)$-transformation that brings the vector into this form
must be of the form
\begin{eqnarray}
 \begin{pmatrix} \beta &  - \alpha \\ \delta & - \gamma \end{pmatrix} 
 \begin{pmatrix} \alpha \\ \beta \end{pmatrix} = \begin{pmatrix} 0 \\ 1 \end{pmatrix}\;,
 \label{trafo2}
\end{eqnarray}
where the integers $(\gamma, \delta)$ are coprime and satisfy
$\alpha \delta - \beta \gamma = 1$,  with $\delta = 0 \mod 2$ and $\gamma$ odd.
This follows by using a similar reasoning as above.

Hence, the vector $V$ can be brought into the form
\begin{eqnarray}
 V = g_V \, \begin{pmatrix} 1   \\ 0 \end{pmatrix}
 \qquad {\rm or} \qquad
 V = g_V \, \begin{pmatrix} 0  \\ 1
 \end{pmatrix}
 \end{eqnarray}
by means of a $\Gamma_0(2)$-transformation.
This result will be used in what follows next.

Now, associate to the  vector $Q^T_m = (q_1, p^0, p^3, p^2)$ the matrix
\begin{eqnarray}
Q \equiv
 \begin{pmatrix} q^1  &  -p^3  \\ p^2  & p^0  \end{pmatrix} \;,
 \label{Qmat}
 \end{eqnarray}
whose determinant equals the charge bilinear $m$, $\det Q =m$.
Let us denote by $\gcd(Q)$ the greatest common divisor of the four charges $(q^1 , p^3 , p^2 , p^0)$.
We now factor out $\gcd(Q)$ from
the matrix $Q$, so that in the following, $Q$ will denote the matrix \eqref{Qmat}
with $\gcd(Q) = 1$.

Now we operate with matrices $A \in \Gamma_0(2)_{U}, \; B \in  \Gamma_0(2)_{T} $ on $Q$ as follows,  $Q \mapsto A Q B^T$, to bring $Q$ either 
into diagonal form or into anti-diagonal form. Note that $\gcd(Q)=1$ is preserved by this operation.
Here, the matrix $A$ takes the form \eqref{gam02}, with the entries $(b,c)$ replaced by $(-b,-c)$,
while the matrix $B$ is of the form \eqref{gam02}.
We proceed to explain this construction.
In the first step, we use the result given above to bring the first column of the matrix $Q$ into the form 
\begin{eqnarray}
Q = 
 \begin{pmatrix} g_V  &  - {\tilde p}^3 \\  0 & {\tilde p}^0  \end{pmatrix}
 \label{Q1}
 \end{eqnarray}
 when $\tilde{p}^2 = 0 \mod 2$, and into the form 
  \begin{eqnarray}
Q = 
 \begin{pmatrix} 0  &  - {\tilde p}^3  \\ g_V  & {\tilde p}^0  \end{pmatrix}
 \label{Q2}
 \end{eqnarray}
 when $\tilde{p}^2 $ is odd.

 Let us first consider the case \eqref{Q1}.  We operate on it from the left with the $\Gamma_0(2)$-matrix
 \begin{eqnarray}\label{k transformation}
 A =   \begin{pmatrix} 1  & k  \\ 0  & 1  \end{pmatrix} \;\;\;,\;\;\; k = \prod_i p_i, \;\;\; \;\;\; p_i \in \{ p | p \, {\rm is \, prime }, p|g_V, \, p \nmid | \tilde{p}^3| \} \;,
 \end{eqnarray}
to obtain
\begin{eqnarray}
 \begin{pmatrix} g_V  &  - {\tilde p}^3 + k {\tilde p}^0 \\  0 & {\tilde p}^0  \end{pmatrix} \;,
 \label{Q3}
 \end{eqnarray}
where we note that the entries of the first row are coprime, for the following reason.  
Let $p$ denote a prime factor of $g_V$. Then, it either divides ${\tilde p}^3$, or it does not.
If it divides ${\tilde p}^3$, it cannot divide $k$, and it also cannot
divide
${\tilde p}^0$, because $\gcd(Q)=1$, by assumption.
If it does not divide ${\tilde p}^3$, it divides $k$. Hence, the two entries of the first row are coprime.

Next, we operate on \eqref{Q3} from the right to bring the first row into canonical form, i.e.
either into the form $(1 \; 0)$ or into the form $(0 \; 1)$,
depending on whether  $- {\tilde p}^3 + k {\tilde p}^0$ is even or odd.
This is achieved by using the transpose of the matrix  given in either \eqref{trafo1}
or \eqref{trafo2}. Therefore, if $- {\tilde p}^3 + k {\tilde p}^0$ is even, we obtain
\begin{eqnarray}
 \begin{pmatrix} 1 &  0
 \\
 r & s 
  \end{pmatrix} \;,
  \label{R1}
 \end{eqnarray}
with some integer entries $r$ and $s$, whereas if $- {\tilde p}^3 + k {\tilde p}^0$ is odd,
we obtain
\begin{eqnarray}
 \begin{pmatrix} 0 &  1
 \\
 {\tilde r} & {\tilde s} 
  \end{pmatrix} \;,
  \label{R2}
 \end{eqnarray}
with some integer entries $\tilde r$ and $\tilde s$. Let us first consider the case \eqref{R1}.
If $r = 0 \mod 2$, we operate on the matrix \eqref{R1} from the left with
\begin{eqnarray}
 \begin{pmatrix} 1 &  0
 \\
-  r & 1 
  \end{pmatrix} \;,
 \end{eqnarray}
to bring the charge matrix into the diagonal form
\begin{eqnarray}
 \begin{pmatrix} 1 &  0
 \\
0 & s
  \end{pmatrix} \;,
 \end{eqnarray}
 with $s = m = \det(Q)$, 
which corresponds to a charge vector $Q^T_m =( 1, m, 0, 0)$.
If $r$ is odd, we operate on the matrix \eqref{R1} from the left with the $\Gamma_0(2)$ matrix
\begin{eqnarray}
 \begin{pmatrix} 1 &  0
 \\
-  (r-1) & 1 
  \end{pmatrix} \;,
 \end{eqnarray}
to obtain
\begin{eqnarray}
 \begin{pmatrix} 1 &  0
 \\
1 & s
  \end{pmatrix} \;.
  \label{mp}
 \end{eqnarray}
 If $s$  is even, we operate on this matrix from the right with the matrix
 \begin{eqnarray}
 \begin{pmatrix} 1 &  -s
 \\
0 & 1
  \end{pmatrix} \;,
 \end{eqnarray}
to obtain
 \begin{eqnarray}
 \begin{pmatrix} 1 & -s 
 \\
1 & 0
  \end{pmatrix} \;.
 \end{eqnarray}
Finally, operating from the left with the matrix
 \begin{eqnarray}
 \begin{pmatrix} 1 & -1
 \\
0 & 1
  \end{pmatrix} \;,
 \end{eqnarray}
we obtain for the charge matrix
 \begin{eqnarray}
 \begin{pmatrix} 0 & -s
 \\
1 & 0
  \end{pmatrix} \;,
 \end{eqnarray}
 with $s = m = \det(Q)$, 
which corresponds to a charge vector $Q^T_m =(0,  0, m, 1)$.
If, on the other hand, $s$ is odd, we operate on \eqref{mp} with
\begin{eqnarray}
\begin{pmatrix} 1 &  0
 \\
s-1 & 1
  \end{pmatrix} 
   \begin{pmatrix} 1 &  0
 \\
1 & s
  \end{pmatrix} 
\begin{pmatrix} 1 &  0
 \\
- 1 & 1
  \end{pmatrix} = \begin{pmatrix} 1 &  0
 \\
0 & s
  \end{pmatrix} \;,
 \end{eqnarray}
 with $s = m = \det(Q)$, 
which corresponds to a charge vector $Q^T_m =(1,  m, 0, 0)$.

Next, let us consider the case \eqref{R2}. If ${\tilde s} = 0 \mod 2$, we operate on \eqref{R2} from the left
with the $\Gamma_0(2)$ matrix
 \begin{eqnarray}
 \begin{pmatrix} 1 & 0
 \\
- {\tilde s} & 1
  \end{pmatrix} \;, 
 \end{eqnarray}
to obtain
 \begin{eqnarray}
 \label{negc}
 \begin{pmatrix} 0 & 1 
 \\
{\tilde r} & 0
  \end{pmatrix} \;, 
 \end{eqnarray}
 with ${\tilde r} = - m = - \det(Q)$, 
which corresponds to a charge vector $Q^T_m =(0,  0, - 1 , -m)$.
By operating with the generator $- \mathbb{I}$ on \eqref {negc} we 
bring this charge vector into the form  $Q^T_m =(0,  0, 1 , m)$.
 If ${\tilde s} $ is odd, we operate on \eqref{R2}
 from the left with the $\Gamma_0(2)$ matrix
\begin{eqnarray}
 \begin{pmatrix} 1 &  0
 \\
-  ( {\tilde s} -1) & 1 
  \end{pmatrix} \;,
 \end{eqnarray}
to obtain
\begin{eqnarray}
 \begin{pmatrix} 0 &  1
 \\
{\tilde r} & 1
  \end{pmatrix} \;.
  \label{tilr2}
 \end{eqnarray}
 Now we 
operate from the right with 
 \begin{eqnarray}
 \begin{pmatrix} 1 &  0
 \\
- {\tilde r} & 1
  \end{pmatrix} 
 \end{eqnarray}
 to obtain
 \begin{eqnarray}
 \begin{pmatrix} - {\tilde r} &  1
 \\
0 & 1
  \end{pmatrix} \;,
   \end{eqnarray}
 and subsequently from the left with 
 \begin{eqnarray}
 \begin{pmatrix} 1 &  - 1
 \\
0 & 1
  \end{pmatrix} \;,
   \end{eqnarray}
to obtain
 \begin{eqnarray}
 \begin{pmatrix} - {\tilde r} & 0
 \\
0 & 1
  \end{pmatrix} \;,
   \end{eqnarray}
 with ${\tilde r} = - m = - \det(Q)$, 
which corresponds to a charge vector $Q^T_m =( m ,  1, 0,0)$.

Finally, let us consider the case when $\tilde{p}^2$ is odd.  Then, we act 
on the left of \eqref{Q2} with the matrix \eqref{k transformation}, to obtain
\begin{equation}\label{Q4}
\begin{pmatrix}
kg_V & -\tilde{p}^3+k\tilde{p}^0\\
g_V & \tilde{p}^0
\end{pmatrix} \;.
\end{equation}
The elements in the first line of \eqref{Q4} are coprime, for the reason given below
\eqref{Q3}, and therefore,
following the same steps as before, we conclude that the case of odd
$\tilde{p}^2$ reduces to the case of even $\tilde{p}^2$.\\

Hence, we have shown that when  $\gcd (q_1, p^0, p^3, p^2) = 1$,
we can bring $Q_m$ into one of the
following
 four inequivalent forms, 
 \begin{eqnarray}
Q^T_m &=&   (1, m, 0, 0) \;\;\;,\;\;\;  m \in \mathbb{Z} \;.
 \nonumber\\
Q^T_m &=&   ( m, 1, 0, 0)  \;\;\;,\;\;\;  m \in \mathbb{Z} \;.
 \nonumber\\
Q^T_m &=&  (0, 0, m, 1)  \;\;\;,\;\;\;  m \in \mathbb{Z} \;.
\nonumber\\
Q^T_m &=&  (0, 0, 1, m) \;\;\;,\;\;\; m \in  \mathbb{Z} \;.
\end{eqnarray}

\end{proof}

%%%%%%%%%%
\section{Small BPS black holes in the STU-model}
%%%%%%%%%

BPS black holes in $N=2$ supergravity theories  carry electric/magnetic charges $(q_I, p^I)$\footnote{The charges $(q_I, p^I)$ used in this paper are those defined in \cite{Cardoso:2000}.}. These charges give rise to three charge
bilinears $(n,m,l)$, which for the STU-model are given by \eqref{bilin}.  Then, one introduces the quartic charge combination
$ \Delta \equiv 4 nm - l^2$, to define small BPS black holes, as follows.\\

\noindent
{\bf Definition:}
A small BPS black hole is a BPS black hole carrying electric/magnetic charges $(q_I, p^I)$ such that the charge combination $\Delta =4 n m - l^2$ vanishes, and such that its macroscopic (Wald) entropy ${\cal S}$ is, for large charges, given by ${\cal S} \propto \sqrt{Q^2}$. Here $Q^2$ denotes a linear combination of
products of two charges.
\\

\noindent
{\it Remark:} At the two-derivative level, a BPS black hole 
has a 
non-vanishing horizon area proportional to $\sqrt{\Delta}$, with $\Delta >0$.
Therefore, a small BPS black hole does not exist as a solution to the equations of motion
of the underlying $N=2$ supergravity theory at the two-derivative level. For a small BPS black hole
to exist, higher-curvature corrections need to be taken into account\footnote{  For a recent discussion on the subtleties in defining small BPS black holes we refer to \cite{Cano:2018}.}. These are encoded in 
$\Omega(Y, \Upsilon)$. \\

For the $N=2$ STU-model
of Sen and Vafa \cite{Sen:1995ff}, 
the Wilsonian function $F(Y, \Upsilon) = F^{(0)} (Y) 
+ 2 i   \, \Omega (Y, \Upsilon)$ takes the form \cite{CdWM},\footnote{
In \cite{CdWM},
the coupling functions $\omega^{(n+1)}(S,T,U) $ were determined by exploiting
the duality symmetries of the STU-model.  These were implemented by adding
to $\Omega (Y, \Upsilon)$ a term $2 \Upsilon \gamma \ln Y^0$.  Since
this term is not part of the Wilsonian effective action, we will not consider
it in the following. \label{footn1}
}
\begin{equation}\label{F}
\begin{split}
F^{(0)}(Y)=&- \frac{Y^1 Y^2 Y^3}{Y^0} \;, \\
\Omega(Y,\Upsilon)=&\Upsilon\sum_{n=0}^\infty \left(\frac{\Upsilon}{(Y^0)^2}\right)^n\omega^{(n+1)}(S,T,U)  \;.
\end{split}
\end{equation}
The gravitational coupling functions $\omega^{(n+1)}(S,T,U)$ were determined in \cite{CdWM}, using 
the duality symmetries of the model, namely invariance under $\Gamma_0(2)_S \times 
\Gamma_0(2)_T \times \Gamma_0(2)_U$ transformations and
under triality (i.e. exchanges of $S, T$ and $U$). The coupling functions $\omega^{(n+1)}(S,T,U)$
are all
non-vanishing when evaluated at a generic point in the $S,T, U$-moduli space, and are determined
in terms of the first gravitational coupling function $\omega^{(1)}(S,T,U)$,
\begin{equation}
\label{omegagenerale}
\omega^{(1)}(S,T,U)=\omega(S)+\omega(T)+\omega(U)\;,
\end{equation}
where \cite{Gregori:1999ns}
\begin{equation}
\omega(S)=
-\frac{\gamma}{2}\ln\vartheta_2^8(iS),\qquad \gamma=-\frac{1}{256\pi},\qquad  \vartheta_2(iS)=2\eta^2(2iS)/\eta(iS).
\end{equation}
Note that
\begin{equation}
{\rm Re} \,S > 0 \;\;\;,\;\;\; {\rm Re} \, T > 0 \;\;\;,\;\;\; {\rm Re} \, U > 0 \;,
\label{reSTUre}
\end{equation}
in view of the fact that $\vartheta_2 (\tau)$ is defined on the complex
upper half-plane $\{\tau \in \mathbb{C}: {\rm Im} \, \tau > 0\}$.

The expressions for the coupling functions $\omega^{(n+1)}(S,T,U)$ obtained in \cite{CdWM} are complicated.
However, as noted in \cite{CdWM}, they do simplify when some of the moduli $S, T, U$ are taken to
large or to small values.  In the following, we will choose a regime where these couplings simplify, namely 
the regime where ${\rm Re} \, S \gg1$ and ${\rm Re} \, T \ll1, {\rm Re} \, U \ll 1$. We proceed with a discussion
of the $\omega^{(n+1)}(S,T,U)$ in this regime.

We begin by discussing the behaviour of $\omega(\tau)$ in the limit of large and of small ${\rm Im} \, \tau$.
Using the product representation for $\vartheta_2 (\tau)$, valid in the punctured disc $\{ q \in \mathbb{C} \backslash \{0\}: |q| <1 \}$, 
\begin{equation}
\vartheta_2 (\tau) = 2\, q^{1/8} \, \prod_{n=1}^{\infty} (1-q^n) \, (1 + q^n)^2 \;\;\;,\;\;\; q= e^{2 \pi i \tau} \;,
\label{exp-theta}
\end{equation}
we obtain for large ${\rm Im} \, \tau \gg 1$, 
\begin{equation}\label{theta2largetau}
\ln \vartheta_2^8 (\tau) =  2 \pi \, i \tau   + {\cal O} (q)  \;,
\end{equation}
and hence, setting $S = -i \tau$,
\begin{equation}
\omega(S) =  \gamma \, \pi \,  S   + {\cal O} (q)  \;,
\end{equation}
For small ${\rm Im} \, \tau \ll 1$, we proceed as follows. Using the relations
\begin{equation}
\vartheta_2(\tau)=\sqrt{\frac{i}{\tau}}\vartheta_4(-\frac{1}{\tau})
\end{equation}
as well as 
\begin{equation}
\vartheta_4(\tau)=\prod_{n=1}^\infty(1-q^n)(1-q^{n-1/2})^2,
\end{equation}
we obtain, setting $T = -i \tau$,
\begin{equation}
\omega(T) = 2 \gamma \, \ln T + {\cal O} (e^{- 2 \pi/T}) \;,
\label{omsmallT}
\end{equation}
provided ${\rm Re} (1/T) \gg 1$. The latter will be enforced in the next subsection by constructing a BPS configuration
that satisfies ${\rm Im} \, T = 0$.

In the regime mentioned above, 
the higher gravitational coupling functions $\omega^{(n+1)}(S,T,U)$, with $n \geq 1$, simplify, and become
proportional to
\begin{equation}
\omega^{(n+1)}(S,T,U) \propto \frac{\gamma^{n+1}}{(T U)^n} \;.
\end{equation}
This is a consequence of the behaviour
$\partial_S \omega \sim 1, \partial_T \omega \sim 1/T, 
\partial_U \omega \sim 1/U$ in this regime, as well as of the explicit form of the $\omega^{(n+1)}(S,T,U)$, with $n \geq 1$, given in \cite{CdWM}.
Then, the function $F(Y, \Upsilon)$ takes the approximate form,
\begin{equation}
F(Y, \Upsilon) = - \frac{Y^1 Y^2 Y^3}{Y^0} + 2 i \Upsilon \left( \gamma \, \pi \, S + 
2\gamma\ln T+ 2\gamma\ln U + \sum_{n =1}^{\infty} 
\left( \frac{\Upsilon }{(Y^0)^2}\right)^n \, \frac{\gamma^{n+1} \, \alpha_n}{(T U)^n} 
\right) \:,
\label{Flim2}
\end{equation}
with constants $\alpha_n \in \mathbb{C}$. Here we assume that the constant coefficients $\alpha_n$ are such that the infinite sum in \eqref{Flim2}, viewed as a power series in $z \equiv 1/[
(Y^0)^2 T U]$, has a non-vanishing
radius of convergence, and that $z = -2$ is in the region of convergence, c.f. \eqref{Y23}.
However, we cannot verify this at the present stage, because of lack of knowledge about the precise form of the constant coefficients
$\alpha_n$. 

In the near-horizon region, a BPS black hole solution is determined in terms of the charges
carried by the black hole. The field $\Upsilon$ takes
a real constant value at the horizon of the BPS black hole, 
and the horizon values of the scalar fields $Y^I$ supporting the
BPS black hole are determined in terms of the charges of the black hole 
by solving the attractor equations \cite{LopesCardoso:1998tkj}
\begin{eqnarray}
Y^I - {\bar Y}^I &=& i p^I \,, \nonumber\\
F_I - {\bar F}_I &=& i q_ I \;, \nonumber\\
\Upsilon &=& - 64 \;,
\label{attrac-eqs}
\end{eqnarray}
where $F_I= \partial F/\partial Y^I$. The semi-classical (Wald) entropy of the BPS black hole
takes the form \cite{LopesCardoso:1998tkj}
\begin{equation}
\mathcal{S}=\pi \left( p^IF_I-q_IY^I+4\Upsilon {\rm Im} \, F_{\Upsilon} \right) \vert_{\rm attractor}\;,
\label{wald}
\end{equation}
where $F_{\Upsilon} = \partial F / \partial \Upsilon$.   
We note the useful relation
\begin{equation}
4 \pi \, \Upsilon \, \gamma = 1 \,.
\label{upsgam1}
\end{equation}

We now proceed to solve the attractor equations \eqref{attrac-eqs} in the regime specified above, with $F(Y,\Upsilon)$ given in \eqref{Flim2}. 
We first construct a small BPS solution with $p^0 \neq 0$ (and magnetic bilinear  $m = 0$). Subsequently, we use the $\Gamma_0(2)_S$ duality 
symmetry of the model
to construct a small BPS black hole with $p^0 =0$ (and $m \neq 0$).

%%%%%%%%%%%%

\subsection{Small BPS black holes with $p^0 \neq 0$ \label{sec:smallp0}}
%%%%%%%%%%%%%%%%%

We construct small BPS black holes with non-vanishing charges $(p^0, p^1, q_2, q_3)$.
The magnetic bilinear $m$ vanishes.

Small BPS black holes with charges $(p^0, q_2, q_3)$
were considered in \cite{Cardoso:2008fr}. However, they were
constructed using a non-holomorphic function $F = F^{(0)} + 2 i \Omega$, with a real $\Omega$.
This came about by implementing the duality symmetries of the model through the inclusion
of non-holomorphic terms. In the present setup, however, the duality symmetries are implemented
by adding a term proportional to $\ln Y^0$, following \cite{CdWM}. We will use this setup to analyze small black holes with four charges $(p^0,p^1,q_2,q_3)$, while also paying attention to the behaviour of the higher gravitational coupling functions $\omega^{(n+1)}(S,T,U)$. The three-charge black hole of \cite{Cardoso:2008fr} will arise as a special case of this four-charge system.

We will work in the regime mentioned above, namely
 large ${\rm Re} \, S \gg 1$, 
 and small  ${\rm Re} \, T \ll1,  \,  {\rm Re} \, U \ll1$. The function $F$ is given in \eqref{Flim2}.
 Introducing $c_1 = - 1/128$, we write $F$ as
 \begin{equation}
\label{F2small}
F=-\frac{Y^1Y^2Y^3}{Y^0}-i\Upsilon c_1\left[i\frac{Y^1}{Y^0}-\frac{2}{\pi}\ln\left(-i\frac{Y^2}{Y^0}\right)-\frac{2}{\pi}\ln\left(-i\frac{Y^3}{Y^0}\right)
+ \sum_{n =1}^{\infty} 
 \frac{\beta_n}{(Y^2 Y^3 )^n} 
\right],
\end{equation}
where 
$\beta_n = (-)^{n+1}  (\Upsilon \gamma)^n \alpha_n/\pi $. 
On the attractor $\Upsilon = -64$, 
the $\beta_n$ are constant, $\beta_n = 4 (-1/4 \pi)^{n+1} \, \alpha_n$,
and $\Upsilon c_1 = 1/2 > 0$.

The magnetic attractor equations \eqref{attrac-eqs} imply that
\begin{equation}
Y^2 = {\bar Y}^2  \;,\; Y^3 = {\bar Y}^3 \;,
\end{equation}
as well as
\begin{equation}
Y^0 = \frac12 \left( \phi^0 + i p^0 \right), \; Y^1=\frac12\left(\phi^1+ip^1\right) \;.
\end{equation}
Using this, the electric attractor equation $F_1 - {\bar F}_1 =0$ gives
\begin{equation}
Y^2Y^3=\Upsilon c_1 \;.
\label{Y23}
\end{equation}
Inserting this into the electric attractor equation $F_0-\bar{F}_0=0$, we get
\begin{equation}\label{phi0null}
\phi^0=0 \;,
\end{equation}
and therefore $Y^0$ is purely imaginary,
\begin{equation}\label{y0*}
Y^0=\frac{ip^0}{2}.
\end{equation}
The electric attractor equation $F_2 - {\bar F}_2 =i q_2$ yields
\begin{equation}
\frac{2\phi^1   \Upsilon c_1}{p^0} + \left( \frac{4  \Upsilon c_1}{\pi} + 2 
\sum_{n =1}^{\infty} 
 \frac{n \, \beta_n}{( \Upsilon c_1 )^n} \right) = q_2 \, Y^2\;,
 \label{Y12}
 \end{equation}
where we used \eqref{Y23}. Similarly, the electric attractor equation 
$F_3 - {\bar F}_3 =i q_3$ yields
\begin{equation}
\frac{2\phi^1   \Upsilon c_1}{p^0} + \left( \frac{4  \Upsilon c_1}{\pi} + 2 
\sum_{n =1}^{\infty} 
 \frac{n \, \beta_n}{( \Upsilon c_1 )^n} \right) = q_3 \, Y^3 \;.
  \label{Y13}
 \end{equation}
Comparing \eqref{Y12} with \eqref{Y13}, we infer
\begin{equation}
q_2 Y^2 = q_3 Y^3 \;.
\label{qYqY}
\end{equation}
Multiplying $Y^2Y^3=\Upsilon c_1 $ with $q_2 q_3$ and using \eqref{qYqY} yields
\begin{equation}
(q_2 Y^2)^2 = q_2 q_3 \Upsilon c_1 > 0 \;.
\end{equation}
Hence we infer
\begin{equation}
q_2 q_3 > 0 
\label{q2q3pos}
\end{equation}
as well as 
\begin{eqnarray}
Y^2 &=& \pm \frac{\sqrt{q_2 q_3 \Upsilon c_1}}{q_2} \;, \nonumber\\
Y^3 &=& \pm \frac{\sqrt{q_2 q_3 \Upsilon c_1}}{q_3} \;.
\label{valueq2Y2Y3}
\end{eqnarray}
Inserting this into \eqref{Y12} or  \eqref{Y13} determines
\begin{equation}
\phi^1 = \frac{p^0}{2 \Upsilon c_1}  \left( - \frac{4 \Upsilon c_1}{\pi}
-  2 
\sum_{n =1}^{\infty} 
 \frac{n \, \beta_n}{( \Upsilon c_1 )^n} 
 \pm \sqrt{q_2 q_3 \Upsilon c_1}  \right) \;.
\label{valueY1}
\end{equation}
Next, we display the expressions for $S, T, U$,
\begin{eqnarray}\label{STU}
S &=& -
\frac{1}{2\Upsilon c_1}  \left( - \frac{4 \Upsilon c_1}{\pi}
-  2 
\sum_{n =1}^{\infty} 
 \frac{n \, \beta_n}{( \Upsilon c_1 )^n} 
 \pm \sqrt{q_2 q_3 \Upsilon c_1}  \right)-i\frac{p^1}{p^0}
\;, \nonumber\\
T &=&  \mp \frac{2 
\sqrt{q_2 q_3 \Upsilon c_1}}{q_2p^0 } \;, \nonumber\\
U&=& \mp \frac{2 
\sqrt{q_2 q_3 \Upsilon c_1}}{q_3p^0 } \; .
\end{eqnarray}
Note that ${\rm Im}\, T = {\rm Im}\, U = 0$, in agreement with what was stated below \eqref{omsmallT}.

Demanding $ {\rm Re} \, S \gg 1$, and recalling that $\Upsilon c_1 = 1/2 > 0$, we infer
that we have to pick the negative sign in the expression for $ {\rm Re} \, S \gg 1$
given in \eqref{STU}, so that
\begin{equation}
\begin{split}
&{\rm Re} S= \frac{1}{2\Upsilon c_1}  \left( \frac{4 \Upsilon c_1}{\pi}+2\sum_{n =1}^{\infty}\frac{n \, \beta_n}{( \Upsilon c_1 )^n}+\sqrt{q_2 q_3 \Upsilon c_1}  \right), \\
&{\rm Re} T=\frac{2\sqrt{q_2 q_3 \Upsilon c_1}}{q_2p^0 }, \\
&{\rm Re} U=\frac{2\sqrt{q_2 q_3 \Upsilon c_1}}{q_3p^0}. 
\end{split}
\end{equation}
Demanding ${\rm Re} \, T > 0, \, {\rm Re} \,U > 0$ requires taking
\begin{equation}\label{qp0 pos}
q_2p^0 > 0 \;\;\;,\;\;\; q_3p^0 > 0 \;,
\end{equation}
and hence, using \eqref{q2q3pos}, we infer that the charges $p^0, q_2, q_3$
have to be either all positive or all negative. To ensure ${\rm Re} \, S \gg 1$, we require $|q_2| \gg1, \, |q_3| \gg 1$.
To ensure
that ${\rm Re} \, T$ and ${\rm Re} \,  U$ are small, we take $|q_2| \sim |q_3|$.
We therefore take 
$|q_2| \gg1, \, |q_3| \gg 1$, $|q_2| \sim |q_3|$, 
$|p^0| \gg1$ to achieve ${\rm Re} \,  T \ll 1, \, {\rm Re} \,  U \ll 1$.
Note that the solution depends crucially on the combination $\Upsilon c_1 \neq 0$, and
hence only exists due to the gravitational coupling functions encoded in 
$\Omega$, as expected for a small BPS black hole.

Next, we compute the semi-classical Wald entropy 
\eqref{wald} of this small BPS black hole, retaining only contributions proportional to charges.
First, we note that on the solution, 
\begin{equation}
F_1=0,
\end{equation}
as well as 
\begin{equation}
p^0 F_0 = - \frac{8 \Upsilon c_1}{\pi} \;,
\end{equation}
which we drop, since it is independent of charges.
Next, using \eqref{qYqY}
as well as \eqref{valueq2Y2Y3} (where we have to take the negative sign), we obtain
\begin{eqnarray}
- q_I Y^I = -2 q_2 Y^2 = 2 \sqrt{q_2 q_3 \Upsilon c_1} \;.
\end{eqnarray}
And finally, using
\begin{equation}
F_{\Upsilon} =  i  c_1 \left[ S 
%+ \frac{2}{\pi}\ln{\left(Y^2 Y^3\right)}
- \frac{4}{\pi}\ln Y^0 
 \right] =  i  c_1 \left[ S - \frac{4}{\pi}\ln |p^0| \right] \;,
 \end{equation}
 which is valid in the approximation that we drop
terms independent of charges, we obtain
\begin{equation}
4 \Upsilon \, \text{Im}F_{\Upsilon} = 4 \Upsilon c_1 \left[{\rm Re}S - \frac{4}{\pi}\ln |p^0| \right] 
= \sqrt{\Upsilon c_1}   \sqrt{q_2 q_3} - 
\frac{ 16 \Upsilon c_1 }{\pi} \ln |p^0| \;,
\end{equation}
again up to a charge independent constant. Hence, for large charges, we obtain,
with $\Upsilon c_1 = 1/2$,
\begin{eqnarray}
\mathcal{S} &=& 4 \pi  \sqrt{\Upsilon c_1}   \sqrt{q_2 q_3} - 16 \Upsilon c_1  \ln |p^0| \nonumber\\
&=& 2 \pi   \sqrt{ 2 \, q_2 q_3} - 8  \ln |p^0|  \;.
\label{waldsmal}
\end{eqnarray}
In the limit where the charges $p^0, q_2, q_3$ are taken to be uniformly large, the first
term is the leading term, and it equals $A/2$, where
$A$ denotes the area of the event horizon of the small BPS black hole.
Note that the result for the BPS entropy \eqref{waldsmal} does not depend on the charge $p^1$.
The term $\ln |p^0| $ represents a correction term, whose coefficient will however
change when including corrections to the semi-classical Wald entropy 
computed by the quantum entropy function \cite{Sen:2008yk,Sen:2008vm,Cardoso:2019avb}
(c.f. footnote \ref{footn1}), and hence
we drop this term in \eqref{waldsmal}. Then, the result \eqref{waldsmal} agrees 
with the one obtained previously in  \cite{Cardoso:2008fr}.

In the above, we have assumed that the coefficients $\alpha_n$ and 
$\beta_n$ are such that the infinite sums in \eqref{F2small} and in \eqref{valueY1} are
convergent.  As already mentioned below \eqref{Flim2}, we cannot verify this at the present
stage.
At any rate, these infinite sums
give rise to charge independent contributions to the semi-classical
entropy, which we discarded. 

When dropping all charge independent terms, the solution 
 given in \eqref{y0*} and \eqref{STU} takes the form 
\begin{eqnarray}
Y^0 &=& \frac{ip^0}{2} \;, \nonumber\\
S &=& 
\frac{1}{2}
  \sqrt{\frac{q_2 q_3}{ \Upsilon c_1}}  -i\frac{p^1}{p^0}
\;, \nonumber\\
T &=&  \frac{2 
\sqrt{q_2 q_3 \Upsilon c_1}}{q_2p^0 } \;, \nonumber\\
U&=& \frac{2 
\sqrt{q_2 q_3 \Upsilon c_1}}{q_3p^0 } \; .
\label{approxsmall}
\end{eqnarray}
We will use this solution to generate a small BPS black hole solution with $p^0=0$
in the next subsection.

%%%%%%%%%%%%%%%%%%%%
\subsection{Small BPS black holes with $p^0 =0$}\label{smallp0=0}
%%%%%%%%%%%%%%%%%

Now we construct small BPS black holes with $p^0 =0$ and non-vanishing magnetic bilinear $m$.
These black holes will have to carry non-vanishing
charges $p^1, p^2, p^3$, to ensure that the real part of the moduli $S, T, U$ is non-vanishing.

In order to obtain a solution with $p^0=0$, we start from the solution 
\eqref{approxsmall} and apply a duality transformation to it, as follows. We act with a 
$\Gamma_0(2)_S$ transformation on the charges, rewriting  explicitly \eqref{QeQmtrasf}:
\begin{equation}\label{trasfcharges}
\begin{split}
p^0 \rightarrow dp^0+cp^1,\qquad  & q_0 \rightarrow aq_0-bq_1, \\
p^1\rightarrow ap^1+bp^0,\qquad  & q_1\rightarrow dq_1-cq_0 , \\
p^2\rightarrow dp^2-cq_3,\qquad  & q_2\rightarrow aq_2-bp^3, \\
p^3\rightarrow dp^3-cq_2,\qquad  & q_3\rightarrow aq_3-bp^2.
\end{split}
\end{equation}
The moduli $Y^I$ transform in a similar manner, by performing the replacement $p^I\to Y^I $, $q_I\to F_I$ in \eqref{trasfcharges}.
We apply the following $\Gamma_0(2)_S$ transformation to \eqref{approxsmall},
\begin{equation}\label{gamma02tr}
\begin{pmatrix}
 a&  & b\\
c& & d \\
\end{pmatrix} = 
\begin{pmatrix}
 1&  & 0\\
-2& & 1 \\
\end{pmatrix} \;.
\end{equation}
To ensure that the resulting charge configuration has vanishing charge $p^0$, we choose the charge $p^1$ in 
\eqref{approxsmall} to have the value 
$p^1 = p^0/2$.  Then, using \eqref{trasfcharges}, we obtain a new charge configuration $(\tilde{q}_I, \tilde{p}^I)$ with 
$\tilde{p}^0 =0$ and with non-vanishing charges $(\tilde{q}_2 = q_2, \tilde{q}_3= q_3, \tilde{p}^1 = p^0/2, \tilde{p}^2 = 2 q_3, \tilde{p}^3=2 q_2)$.
The magnetic charge bilinear $m$ is non-vanishing and given by $m = \tilde{p}^2 \tilde{p}^3 = 4 q_2 q_3$.

The associated moduli $\tilde{Y}^I$ are, to leading order in the charges, given by
\begin{equation}\label{transfmoduli}
\begin{split}
&\tilde{Y}^0= Y^0 - 2 Y^1 = \frac{p^0}{2} \sqrt{ \frac{q_2q_3}{\Upsilon c_1} } 
,\\
&\tilde{Y}^1= Y^1 = -\frac{p^0}{4}   \sqrt{ \frac{q_2q_3}{\Upsilon c_1} } +\frac{ip^0}{4},\\
&\tilde{Y}^2= Y^2 + 2 F_3 =  iq_3 ,\\
&\tilde{Y}^3=Y^3 + 2 F_2 = iq_2 .
\end{split}
\end{equation}
In obtaining these leading order expressions, we used $F = F^{(0)} = - Y^1 Y^2 Y^3/Y^0$, rather than \eqref{F2small}.
These expressions receive subleading corrections in the charges which we have dropped.
Since $\Gamma_0(2)_S$ constitutes a symmetry of the model, the configuration \eqref{transfmoduli}
describes a small BPS black hole solution, to leading order in the charges.

The constraints \eqref{q2q3pos}, \eqref{qp0 pos} on the charges ensure that
the transformed fields satisfy
 ${\rm Re}\, \tilde{S},{\rm Re}\, \tilde{T}, {\rm Re}\, \tilde{U}>0$. 
 In addition, using $|q_2|, |q_3|, |p^0| \gg1, |q_2| \sim |q_3|$ (c.f. below \eqref{qp0 pos}),
 we infer from  \eqref{transfmoduli} 
 that ${\rm Re}\, S,{\rm Re}\, T, {\rm Re}\, U\ll 1$.

%%%%%%%%%%%
%%%%%%%%%%%%
\section{Two-center bound state of small BPS black holes \label{multicentered} }
%%%%%%%%%%%%%%

Using the small BPS black holes constructed in the previous section, we now
construct a two-center bound state of small BPS black holes in asymptotically flat space-time.
In doing so, we take the total charge of this configuration to be the one of a large dyonic BPS black hole.
In addition, we take 
the asymptotic values
of the moduli $Y^I$ to equal the attractor values of the associated dyonic BPS black hole.

Let us denote the charge configuration of a BPS black hole by
\begin{equation}
\begin{pmatrix}
Q_e^T\\
Q_m^T
\end{pmatrix}=
\begin{pmatrix}
q_0 & -p^1 & q_2 & q_3\\
q_1 & p^0 & p^3 & p^2
\end{pmatrix}, 
\end{equation}
with $Q_e^T, Q_m^T$ as in \eqref{chargenotation}. We consider a 
two-center configuration of small BPS black holes with charge vectors
$\Gamma$ and $\tilde\Gamma$ given by
\begin{equation}
\begin{split}
&\Gamma=
\begin{pmatrix}
0 & -p^1 & q_2 & q_3\\
0 & 0 & 2q_2 & 2q_3
\end{pmatrix}, \\
&\tilde\Gamma=
\begin{pmatrix}
0 & -2\tilde{q}_3 & 0 & \tilde{q}_3\\
\tilde{q}_1 & 0 & 2\tilde{q}_1 & \tilde{p}^2
\end{pmatrix} .\\
\end{split}
\end{equation}
Here, the charge vector $\Gamma$ describes the small BPS black hole given in \eqref{transfmoduli}, while
the charge vector 
$\tilde{\Gamma}$ describes another small BPS black hole, obtained from the first one by triality, as follows.
Starting from the small BPS black hole with non-vanishing charges $(p^0, p^1, q_2, q_3)$ constructed
in subsection \ref{sec:smallp0}, we construct a small BPS black hole with non-vanishing charges $(p^0, p^2, q_1, q_3)$ 
by interchanging $I=1 \leftrightarrow I=2$. Then, we apply a $\Gamma_0(2)_T$ transformation to this charge
configuration. Such a transformation acts as in \eqref{trasfcharges} on the charges, with 
$I=1$ and $I=2$ interchanged.
Then, picking the transformation \eqref{gamma02tr}, we obtain $\tilde{\Gamma}$.
Note that $\Gamma$ and $\tilde{\Gamma}$ describe small BPS black holes
whose attractor values $S, T, U$ satisfy ${\rm Re}\, S,{\rm Re}\, T, {\rm Re}\, U\ll 1$. Also, we recall
that the charges of these small BPS black holes are constrained, as discussed below \eqref{qp0 pos}.
For concreteness, we take
\begin{equation}
\label{charconssmall}
p^1, q_2, q_3, \tilde{q}_1, \tilde{q}_3, \tilde{p}^2 > 0 \;.
\end{equation}
The total charge $\Gamma_{\rm tot}= \Gamma + \tilde{\Gamma}$ of the two-center configuration is
\begin{equation}\label{totalcharge}
\Gamma_{\rm tot}=
\begin{pmatrix}
0 & -(p^1+2\tilde{q}_3) & q_2 & q_3+\tilde{q}_3\\
\tilde{q}_1 & 0 & 2(\tilde{q}_1+q_2) & \;  \tilde{p}^2+2q_3
\end{pmatrix} .
\end{equation}
Next, defining the $\Gamma_0(2)_S \times \Gamma_0(2)_T \times \Gamma_0(2)_U$ invariant product
\begin{equation}
\langle\Gamma,\tilde\Gamma\rangle=Q_e^T\eta\tilde{Q}_m-\tilde{Q}_e^T\eta Q_m \;,
\end{equation}
we obtain for the two-center configuration,
\begin{equation}\label{pairing}
\langle\Gamma,\tilde\Gamma\rangle=q_2\tilde{p}^2+2\tilde{q}_1q_3-\tilde{q}_1p^1-2\tilde{q}_3q_2 \;.
\end{equation}

The two-center configuration represents a two-center bound state, provided the 
separation
distance $|x|$ between the two BPS centers is finite and positive \cite{Denef:2000nb,Bates:2003vx}. This distance is given by \cite{Denef:2000nb,Bates:2003vx}
\begin{equation}\label{raggio}
\vert x\vert= \frac12 
\langle \Gamma,\tilde\Gamma\rangle 
\frac{ \vert Z(\Gamma_{\rm tot})\vert}{ {\rm Im} (Z(\Gamma)\bar{ Z}(\tilde{\Gamma}))  }\bigg\vert_{r=\infty} >0 \;,
\end{equation}
where $Z = Z(\Gamma)$ denotes the central charge computed from the prepotential
function $F^{(0)}$ at spatial infinity, 
\begin{equation}
Z(p^I,q_I)=p^I F^{(0)}_I-q_IY^I \;.
\end{equation}
Here, the $Y^I$ denote the values at spatial infinity, which are arbitrary.
In the following, we choose these values to equal the attractor values computed in the near-horizon
region of  a large BPS black hole with total charge \eqref{totalcharge}.
We determine these values by solving the attractor equations \eqref{attrac-eqs} with $ F = F^{(0)}$,
\begin{equation}\label{attr-espl}
\begin{cases}
Y^0=\frac{\phi^0}{2}\\
Y^1=\frac{\phi^1+i(p^1+2\tilde{q}_3)}{2}\\
Y^2=\frac{\phi^2+i(2q_3+\tilde{p}^2)}{2}\\
Y^3=\frac{\phi^3+i(2q_2+2\tilde{q}_1)}{2}
\end{cases}\quad
\begin{cases}
\frac{Y^1Y^2Y^3}{(Y^0)^2}-\frac{\bar{Y}^1\bar{Y}^2\bar{Y}^3}{(\bar{Y}^0)^2}=0\\
-\frac{Y^2Y^3}{(Y^0)}+\frac{\bar{Y}^2\bar{Y}^3}{(\bar{Y}^0)}=i\tilde{q}_1\\
-\frac{Y^1Y^3}{(Y^0)}+\frac{\bar{Y}^1\bar{Y}^3}{(\bar{Y}^0)}=iq_2\\
-\frac{Y^1Y^2}{(Y^0)}+\frac{\bar{Y}^1\bar{Y}^2}{(\bar{Y}^0)}=i(q_3+\tilde{q}_3)\\
\end{cases}
\end{equation}
We obtain the attractor values
\begin{small}

\begin{equation}
\label{fi0123}
\begin{split}
\phi^0&=\pm \frac{4(\tilde{q}_1+q_2)(\tilde{p}^2+2q_3)(p^1+2\tilde{q}_3)}{\sqrt{-(p^1)^2\tilde{q}_1^2-(\tilde{p}^2)^2q_2^2-4(\tilde{q}_1q_3-q_2\tilde{q}_3)^2+4\tilde{p}^2q_2(\tilde{q}_1q_3+2\tilde{q}_1q_3+q_2\tilde{q}_3)+2p^1\tilde{q}_1(2\tilde{q}_1q_3+q_2(\tilde{p}^2+4q_3+2\tilde{q}_3))}} , \\
S&=i\frac{-p^1\tilde{q}_1+\tilde{p}^2q_2+2\tilde{q}_1q_3+4q_2q_3+2q_2\tilde{q}_3}{4(\tilde{q}_1+q_2)(\tilde{p}^2+2q_3)}+\frac{p^1+2\tilde{q}_3}{\phi^0} , \\
T&=i\frac{p^1\tilde{q}_1-\tilde{p}^2q_2+2\tilde{q}_1q_3+4\tilde{q}_1\tilde{q}_3+2q_2\tilde{q}_3}{4(\tilde{q}_1+q_2)({p}^1+2\tilde{q}_3)}+\frac{\tilde{p}^2+2q_3}{\phi^0} , \\
U&=i\frac{p^1\tilde{q}_1+\tilde{p}^2q_2-2(\tilde{q}_1q_3+2q_2\tilde{q}_3)}{2(\tilde{p}^2+2q_3)({p}^1+2\tilde{q}_3)}+\frac{2q_2+2q_3}{\phi^0}.
\end{split}
\end{equation}
\end{small}
This requires to take
\begin{equation}
-(p^1)^2\tilde{q}_1^2-(\tilde{p}^2)^2q_2^2-4(\tilde{q}_1q_3-q_2\tilde{q}_3)^2+4\tilde{p}^2q_2(\tilde{q}_1q_3+2\tilde{q}_1q_3+q_2\tilde{q}_3)+2p^1\tilde{q}_1(2\tilde{q}_1q_3+q_2(\tilde{p}^2+4q_3+2\tilde{q}_3))>0 , 
\label{condsqrt}
\end{equation}
to ensure that the expression for $\phi^0$ is well defined. In addition, we demand
\begin{equation}\label{ReSTU}
\begin{split}
&\text{Re}S=\frac{p^1+2\tilde{q}_3}{\phi^0}>0, \\
&\text{Re}T=\frac{\tilde{p}^2+2q_3}{\phi^0}>0, \\
&\text{Re}U=\frac{2q_2+2q_3}{\phi^0}>0.
\end{split}
\end{equation}

Next, we evaluate \eqref{raggio}, using
\begin{eqnarray}
Z(\Gamma) &=& p^1F_1+2q_3F_2+2q_2F_3-q_2Y^2-q_3Y^3 \;, \nonumber\\
{Z}(\tilde{\Gamma}) &=& 2\tilde{q}_3 {F}_1+\tilde{p}^2 {F}_2+2\tilde{q}_1 {F}_3-\tilde{q}_1Y^1-\tilde{q}_3Y^3 \;.
\end{eqnarray}
For simplicity, we choose the charges \eqref{charconssmall} as follows,
\begin{equation}
 \tilde{p}^2=2p^1= 2 p > 0 \;\;\;,\;\;\; 
q_2 = q_3 = \tilde{q}_1 = \tilde{q}_3 = q > p > 0  \;.
\label{choicechar}
\end{equation}
This ensures that \eqref{condsqrt} is satisfied, and so is \eqref{ReSTU}, provided we choose $\phi^0 >0$ in \eqref{fi0123}.
For  \eqref{pairing} we obtain
\begin{equation}
\langle\Gamma,\tilde{\Gamma}\rangle=p q > 0 \;.
\end{equation}
And finally, we compute
\begin{eqnarray}
\text{Im}(Z(\Gamma)\bar{Z}(\tilde\Gamma))= \frac{p q^2}{\phi^0 (48 q -p) } (3 p^3 - 114 p^2 q - 1780 p q^2 + 640 q^3) \;,
\end{eqnarray}
which is positive for $q > 3 p $. Hence we conclude that for the choice of charges \eqref{choicechar}, with $q > 3 p $,
 the separation \eqref{raggio} is finite and positive.  Thus, the two-center configuration with total charge \eqref{totalcharge}
 describes a two-center bound state of small BPS black holes in asymptotically flat space-time,
 with the asymptotic moduli taken to equal
the attractor values of a dyonic BPS black hole with total charge \eqref{totalcharge}. Note that this bound state is stabilized by a non-vanishing angular momentum given in \eqref{pairing}.

%%%%%%%%%%%%%%%%%%%%%%%%%%%%%%%%%%
\subsection*{Acknowledgements}
I would like to thank G. L. Cardoso and S. Nampuri 
for valuable discussions. This work was 
supported by 
 Funda\c{c}\~{a}o para a Ci\^{e}ncia e Tecnologia (FCT), Portugal,
 through a Lisbon Mathematics PhD (LisMath) fellowship, 
PD/BD/113626/2015.
 %%%%%%%%%%%%%%%%%%%%%%%%%%

\providecommand{\href}[2]{#2}\begingroup\raggedright\endgroup

\end{document}